\documentclass[12pt]{article}
\usepackage[english]{babel}
\usepackage{amsfonts}
\usepackage{amsmath}
\usepackage{amssymb}
\usepackage{mathrsfs}
\usepackage{multicol}
\usepackage{fancybox}
\usepackage{graphicx}
\usepackage{dsfont}
\usepackage{color}
\usepackage{hyperref}
\usepackage{subfigure} 
\usepackage{graphicx} 

\def\tl{\tilde} 
\def\sg{\sigma} 
\def\gm{\gamma} 
\def\lm{\lambda}

\def\be{\begin{equation}}
\def\ee{\end{equation}}
\def\bea{\begin{eqnarray}}
\def\eea{\end{eqnarray}}

\def\dblone{\hbox{$1\hskip -1.2pt\vrule depth 0pt height 1.6ex width 0.7pt \vrule depth 0pt height 0.3pt width 0.12em$}}

\begin{document}

 \bigskip

\begin{center}
{\Large{\bf{B\"acklund Transformations as exact integrable time-discretizations for the trigonometric Gaudin model}}}
\end{center}
\bigskip

\begin{center}
{ {\bf Orlando Ragnisco,  Federico Zullo}}

{Dipartimento di Fisica,   Universit\`a di Roma Tre \\ Istituto Nazionale di
Fisica Nucleare, sezione di Roma Tre\\  Via Vasca Navale 84,  00146 Roma, Italy  \\
~~E-mail: ragnisco@fis.uniroma3.it, zullo@fis.uniroma3.it}

\end{center}

\medskip
\medskip

\begin{abstract}
\noindent
We construct a two-parameter family of  B\"acklund transformations for the trigonometric classical
Gaudin magnet. The approach follows closely the one introduced by E.Sklyanin
and V.Kuznetsov (1998,1999) in a number of seminal papers, and  takes
advantage of the intimate relation between the trigonometric and the rational
case. As in the paper by A.Hone, V.Kuznetsov and one of the authors (O.R.)
(2001) the B\"acklund transformations are presented as explicit symplectic
maps, starting from their Lax representation.  The (expected)  connection with
the \emph{xxz} Heisenberg chain is established and the rational (\emph{xxx}) case is recovered in
a suitable limit. It is shown how to obtain  a ``physical'' transformation
mapping real variables into real variables. The interpolating Hamiltonian flow
is derived and some numerical iterations of the map are presented. 
 \end{abstract}

\bigskip\bigskip\bigskip\bigskip

\noindent


\newpage

\section{Introduction} 
B\"acklund transformations are nowadays a widespread useful tool
related to the theory of nonlinear differential equations. The first
historical evidence of their mathematical significance was given by Bianchi
\cite{Bianchi} and B\"acklund \cite{Backlund} on their works on surfaces of
constant curvature.  A simple approach to understand their
importance can be to regard them as a mechanism allowing to endow a
given nonlinear differential equation with a nonlinear superposition principle
yielding  a set of solutions through a merely \emph{algebraic procedure}
\cite{Rogers},\cite{Adler},\cite{Levi}.  B\"acklund transformations are indeed parametric families of difference equations encoding  the whole set of symmetries of a given integrable dynamical system.  For finite-dimensional integrable  systems the technique of B\"acklund transformations leads to the construction of integrable Poisson maps that discretize a family of continuous flows
\cite{SW},\cite{Ves},\cite{Sur2},\cite{S1},\cite{SK},\cite{KV}. Actually in
the last two decades numerous results have appeared in the field of exact
discretization of many-body integrable systems employing the B\"acklund
transformations tools
\cite{Rag_Sur},\cite{Sur2},\cite{Rag},\cite{KV},\cite{SK},\cite{Nij},\cite{S1}. 
For the \emph{rational} Gaudin model such
discretization has been obtained ten years ago in \cite{HKR}; afterwards,  these
results have been used for constructing an integrable discretization of classical
dynamical systems  (as the Lagrange top) connected to Gaudin model through In\"onu-Wigner
contractions \cite{MPR},\cite{KPR},\cite{MPRS}. 

\noindent
The aim of the present  work is to construct B\"acklund
transformations for the Gaudin model in the partially anisotropic  ($xxz$) case,
i.e. for the \emph{trigonometric} Gaudin model. We point out that  partial results on this issue have already been
given  in \cite{noi}. 

\noindent
The paper is organized as follows. 
\newline
In Section (\ref{sec1}) we review the main features of the trigonometric Gaudin model from   the point of view of its
integrability  structure. For the sake of completeness,  in Section
(\ref{sec2}) we briefly recall the preliminary results on B\"acklund Transformations (BTs)  for trigonometric Gaudin given in \cite{noi}. In Section (\ref{sec3}) the \emph{explicit form} of BTs
is  given; it is shown that they are indeed  a  trigonometric
generalization of the rational ones (see \cite{HKR}) which can be recovered in a suitable (``small angle" ) limit. The simplecticity of the
transformations is also discussed  in the same  Section and the proof allows us to
elucidate the (expected) link between the Darboux-dressing matrix and the
elementary Lax matrix for the $xxz$ Heisenberg magnet on the lattice. We end the
Section by mentioning   an  open  question, namely  the construction of an explicit  generating function for 
these B\"acklund transformations. In Section (\ref{sec4}) we will show how our
map can lead, with an appropriate choice of B\"acklund parameters, to physical
transformations, i.e. transformations from real variables to real
variables. In the last Section we show how  a suitable  continuous limit yields the interpolating Hamiltonian flow and finally present   numerical examples of iteration of the map.  
 
\section{Gaudin magnet in the trigonometric case} \label{sec1}
For a full account of the integrability structure of the classical and quantum Gaudin model we refer the reader to  the  fundamental contributions by Semenov-Tian-Shanski \cite{STS} and Babelon-Bernard-Talon \cite{BT}. In this section we briefly recall the main features of  the trigonometric Gaudin magnet. 
\newline
The Lax matrix of the model is given by the expression:  
\begin{equation}\label{eq:lax} 
L(\lm) =   \left( \begin{array}{cc} A(\lm) & B(\lm)\\C(\lm)&-A(\lm)\end{array} 
\right) 
\end{equation}
\begin{equation} 
\label{ABC} 
A(\lm)=\sum_{j=1}^{N}\cot(\lm-\lm_{j})s^{3}_{j}, \qquad 
B(\lm)=\sum_{j=1}^{N}\frac{s^{-}_{j}}{\sin(\lm-\lm_{j})},\qquad C(\lm)=\sum_{j=1}^{N}\frac{s^{+}_{j}}{\sin(\lm-\lm_{j})}. 
\end{equation} 
In (\ref{eq:lax}) and (\ref{ABC}) $\lm \in \mathbb{C}$ is the spectral parameter, $\lm_{j}$
are arbitrary real parameters of the model, while
$\big(s^{+}_{j},s^{-}_{j},s^{3}_{j}\big)$,\, $j=1, \ldots, N$, are the dynamical
variables of the system obeying to $\oplus^{N} sl(2)$ algebra, i.e.   
\begin{gather} \label{poisS}
\big\{s^{3}_{j},s^{\pm}_{k}\big\}=\mp i\delta_{jk}s^{\pm}_{k}, \qquad 
\big\{s^{+}_{j},s^{-}_{k}\big\}=-2i\delta_{jk}s^{3}_{k}, 
\end{gather} 
By fixing the $N$ Casimirs $ \big(s_{j}^{3}\big)^{2}+s_{j}^{+}s_{j}^{-}\doteq s_{j}^{2}$ one obtains a symplectic manifold given by  the
direct sum  of the correspondent $N$ two-spheres. 
\\
 Reformulating  the  Poisson structure in terms of the $r$-matrix formalism amounts
to state that the Lax matrix satisfies  the \emph{linear} $r$-matrix Poisson algebra (see again \cite{STS}, \cite{BT}) :  
\begin{gather} \label{eq:pois} 
\big\{ L(\lm)\otimes \dblone, \dblone\otimes L(\mu)\big\}=\big[r_{t}(\lm-\mu), L(\lm)\otimes \dblone + \dblone\otimes L(\mu) \big], 
\end{gather} 
where $r_{t}(\lm)$ stands for the trigonometric $r$ matrix \cite{FT}:  
\begin{gather} 
r_{t}(\lm) = \frac{i}{\sin(\lm)}\left(\begin{array}{cccc} \cos(\lm)& 0 & 0 &0 \\ 
0 & 0 & 1 & 0\\ 
0 & 1 & 0 & 0\\ 
0 & 0 & 0 & \cos(\lm) 
\end{array}\right), 
\end{gather} 
Equation (\ref{eq:pois}) entails   the following  Poisson brackets for the functions (\ref{ABC}): 
\begin{gather} 
 \{A(\lm),A(\mu)\}=\{B(\lm),B(\mu)\}=\{C(\lm),C(\mu)\}=0,\nonumber\\ 
 \{A(\lm),B(\mu)\}=i\frac{\cos(\lm-\mu)B(\mu)-B(\lm)}{\sin(\lm-\mu)},\nonumber\\ 
 \{A(\lm),C(\mu)\}=i\frac{C(\lm)-\cos(\lm-\mu)C(\mu)}{\sin(\lm-\mu)},\nonumber\\ 
 \{B(\lm),C(\mu)\}=i\frac{2(A(\mu)-A(\lm))}{\sin(\lm-\mu)}.
\end{gather}
The determinant of the Lax matrix is the generating function of the integrals
of motion:
\begin{equation}\label{generfun}
-\textrm{det}(L)=A^{2}(\lambda)+B(\lambda)C(\lambda)=\sum_{i=1}^{N}\left(\frac{s_{i}^{2}}{\sin^{2}(\lambda-\lambda_{i})}+H_{i}\cot(\lambda-\lambda_{i})\right)-H_{0}^{2}
\end{equation}
where the $N$ Hamiltonians $H_{i}$ are of the form:
\begin{equation} \label{hams}
H_{i}=\sum_{k\neq i}^{N} \frac{2\cos(\lm_{i}-\lm_{k})s_{i}^{3}s_{k}^{3}+ s_{i}^{+}s_{k}^{-}+ s_{i}^{-}s_{k}^{+}}{\sin(\lm_{i}-\lm_{k})} 
\end{equation}
Note that only $N-1$ among these Hamiltonians are independent, because of
$\sum_{i}H_{i}=0$. Another integral is given by $H_{0}$, the projection of the total spin on the $z$ axis: 
\begin{equation} 
H_{0}=\sum_{j=1}^{N}s_{j}^{3} \doteq J^{3} 
\end{equation}  
The Hamiltonians $H_{i}$ are in involution for the Poisson bracket
(\ref{poisS}):
\begin{equation}
\{H_{i},H_{j}\}=0 \quad i,j=0,\ldots ,N-1
\end{equation}
The  corresponding Hamiltonian flows  are then given by: 
\begin{equation} 
\frac{ds^{3}_{j}}{dt_{i}}=\{H_{i},s^{3}_{j}\}         \qquad  
\frac{ds^{\pm}_{j}}{dt_{i}}=\{H_{i},s^{\pm}_{j}\} 
\end{equation}
In the $xxx$ model a remarkable Hamiltonian is found by taking
a linear combination of the integrals corresponding to (\ref{hams}) in the
rational case \cite{FM}. It describes a mean field spin-spin interaction:
$$
\mathcal{H}_{r}=\frac{1}{2}\sum_{i\neq j}^{N}\mathbf{s}_{i}\cdot\mathbf{s}_{j}
$$   
Where the notation for the bold symbol $\mathbf{s}_{i}$ is
$\mathbf{s}_{i}=(s_{i}^{1},s_{i}^{2},s_{i}^{3})$ with
$s_{i}^{+}=s_{i}^{1}+is_{i}^{2}$ and $s_{i}^{-}=s_{i}^{1}-is_{i}^{2}$. The
natural trigonometric generalization of this Hamiltonian can be found by taking the linear combination of \ref{hams}:
$$
\sum_{i=1}^{N}\frac{\sin(2\lambda_{i})}{2}H_{i}
$$
giving
\begin{equation}
\mathcal{H}_{t}=\frac{1}{2}\sum_{i\neq j}^{N}\cos(\lambda_{i}+\lambda_{j})\left(s_{i}^{1}s_{j}^{1}+s_{i}^{2}s_{j}^{2}+\cos(\lambda_{i}-\lambda_{j})s_{i}^{3}s_{j}^{3}\right)
\end{equation} 
\section{A first approach to Darboux-dressing matrix}\label{sec2} 
In this Section, for the sake of  completeness,  we recall the results already appeared in \cite{noi}. The leading observation is that by performing the ``uniformization'' mapping: 
\begin{gather*} 
\lm \to z\doteq e^{i\lm} \label{uni} 
\end{gather*} 
the $N$-sites trigonometric Lax matrix takes a rational form in $z$ that
corresponds to the $2N$-sites rational Lax matrix plus an additional reflection
symmetry (see also \cite{Hik}); in fact, by performing the substitution (\ref{uni}), the Lax matrix (\ref{eq:lax}) becomes: 

\begin{gather} 
\label{eq:ratiL} 
L(z)=iJ^{3}+\sum_{j=1}^{N} \left(\frac{L_{1}^{j}}{z-z_{j}}-\sigma_3\frac{L_{1}^{j}}{z+z_{j}}\sigma_3\right), 
\end{gather} 
where $\sigma_{3}$ is the Pauli matrix $diag(1,-1)$ and  the matrices $L_{1}^{j}$, \, $j=1, \ldots, N$, are given by:   

\begin{gather*} 
L_{1}^{j}=iz_{j}\left(\begin{array}{cc}s_{j}^{3} & s_{j}^{-}\vspace{1mm}\\ 
s_{j}^{+} & -s_{j}^{3}\end{array}\right) 
\end{gather*} 
So, equation (\ref{eq:ratiL}) entails  the following involution on $L(z)$:  
\begin{gather} 
L (z) = \sigma_3 L (-z) \sigma_3\label{Symm} 
\end{gather}
Constructing a B\"acklund transformation for the Trigonometric Gaudin System (TGS) amounts to build up a Poisson map for the field variables of the model (\ref{ABC}) such that the integrals of motion (\ref{hams}) are preserved. At the level of Lax matrices, this transformation is usually seeked as a similarity transformation between an {\it{old}}, or ``undressed'', Lax matrix $L$, and a {\it{new}}, or ``dressed'' one, say $\tilde{L}$:
\begin{equation}\label{eq:invpre} 
L(z)\to D(z)L(z)D^{-1}(z)\equiv \tl{L}(z) 
\end{equation}   
But $L$ and $\tilde{L}$ have to enjoy the same reflection symmetry
(\ref{Symm}) too: to preserve this involution the Darboux dressing matrix $D$
has to share with $L$ the property (\ref{Symm}); the elementary dressing
matrix $D$ is then obtained by requiring the existence of only one pair of
opposite poles for $D$ in the complex plane of the spectral parameter. We will show in the next
Section that, thanks to  this constraint, one recovers the form of the Lax matrix
for the elementary
$xxz$ Heisenberg spin chain: on the other hand, this is quite natural  if one recalls that 
for the rational Gaudin model the elementary Darboux-dressing matrix is given
by the Lax matrix for the elementary $xxx$ Heisenberg spin chain \cite{HKR},\cite{SK}. The previous observations lead
to the following Darboux matrix:
\begin{equation}\label{eq:Ddiz} 
D(z)=D_{\infty}+\frac{D_{1}}{z-\xi}-\sg_{3}\frac{D_{1}}{z+\xi}\sg_{3}  
\end{equation} 
By taking the limit $z \to \infty$ in (\ref{eq:Ddiz}) it is readily seen
that $D_{\infty}$ has to be a diagonal matrix. In order to ensure that $L$ and
$\tilde{L}$ have the same rational structure in $z$, we rewrite equation (\ref{eq:invpre}) in the form: 
\begin{gather} 
\tilde{L} (z) D(z) = D(z) L(z)\label{BTnew} 
\end{gather} 
Now it is clear that both sides have the same residues at the poles $z=
z_{j}$, $z= \xi_{j}$ (it is unnecessary to look at the poles in $z= -z_{j}$
and $z= -\xi_{j}$ because of the symmetry (\ref{Symm}), so that the following set of equations have to be satisfied: 
\begin{gather} 
\tilde L_{1}^{(j)} D(z_{j}) = D(z_{j}) L_{1}^{(j)}, \label{resj} 
\\ 
\tilde L (\xi) D_{1} = D_{1} L(\xi). \label {resxi} 
\end{gather} 
In principle, equations (\ref{resj}), (\ref{resxi}) yield a Darboux matrix
depending \emph{both} on the old (untilded) variables and the new (tilded)
ones, implying in turn an implicit relationship between the same variables. To
get an explicit relationship one has to resort to the so-called spectrality
property \cite{SK} \cite{KV}. To this aim we need to force the determinant of
the Darboux matrix $D(z)$ to have, besides the pair of poles at $z=\pm \xi$, a
pair of opposite \emph{nondynamical} zeroes, say at $z=\pm \eta$, and to allow
the matrix $D_{1}$ to be proportional to a projector \cite{noi}. Again by symmetry it suffices to consider just one of these zeroes.  
If $\eta$ is a zero of det$D(z)$, then $D(\eta)$ is a rank one matrix, possessing a one dimensional kernel $|K(\eta)\rangle$;  the equation (\ref{BTnew}) : 
\begin{gather} 
\tilde{L} (\eta) D(\eta) = D(\eta) L(\eta) 
\end{gather} 
entails 
\begin{gather} 
D(\eta) L(\eta)|K(\eta)\rangle =0. 
\end{gather} 
This equation in turn allows to infer that $|K(\eta)\rangle$ is an eigenvector for the Lax matrix $L(\eta)$: 
\begin{gather} 
\label{spectral1} 
 L(\eta)|K(\eta)\rangle = \mu (\eta) |K(\eta)\rangle, 
\end{gather} 
This relations gives a direct link between the parameters appearing in the
dressing matrix $D$ and the \emph{old} dynamical variables in $L$. Because of (\ref{resxi}) we have another one dimensional kernel $|K(\xi)\rangle$ of $D_{1}$, such that: 
\begin{gather} 
\label{spectral2} 
 L(\xi)|K(\xi)\rangle = \mu (\xi) |K(\xi)\rangle. 
\end{gather} 
In \cite{noi} we have shown how the two spectrality conditions (\ref{spectral1}), (\ref{spectral2}) enable to 
write $D$ in terms of the old dynamical variables and of the two B\"acklund 
parameters $\xi$ and $\eta$. The explicit expression of the Darboux dressing matrix is given by: 
 
\begin{gather} 
\label{Dexpl} 
D(z)= \frac{\beta z}{z^{2}-\xi^{2}}\left( \begin{array}{cc} {\frac {z \left( p(\eta)\eta-p(\xi)\xi \right) 
    }{b}}+{\frac{ \left( p(\xi)\eta-p(\eta)\xi \right) \eta\xi}{b z}}&{\frac {{\xi}^{2}-{\eta}^{2}}{b}}\vspace{2mm}\\ 
 {\frac {b p(\xi)p(\eta) \left( 
      {\xi}^{2}-{\eta}^{2} \right) }{\eta\xi}}&{\frac {b \left( p(\eta)\eta-p(\xi)\xi \right) 
    }{z}}+{\frac {  b z\left( p(\xi)\eta-p(\eta)\xi \right)}{\eta\xi}}\end{array} \right). 
\end{gather} 
In this expression $\beta$ is a global multiplicative factor, inessential with respect to the form of the
BT, $b$ is an
undeterminate parameter that in Section (\ref{sec3}) we will fix in
order to recover the form of the Lax matrix for the discrete $xxz$ Heisenberg spin chain. The functions
$p(\eta)$ and $p(\xi)$ characterize completely the kernels of $D(\eta)$ and
$D(\xi)$: in fact we have the following formulas \cite{noi}: 
\begin{equation} 
|K(\xi)\rangle=\left(\begin{array}{c}1\\p(\xi)\end{array}\right) \qquad |K(\eta)\rangle=\left(\begin{array}{c}1\\p(\eta)\end{array}\right) 
\end{equation}  
As $|K(\xi)\rangle$ and $|K(\eta)\rangle$ are respectively eigenvectors of $L(\xi)$ and $L(\eta)$, $p(\xi)$ and $p(\eta)$ 
must satisfy: 
\begin{gather} 
p(\xi)=\frac{\mu(\xi)-A(\xi)}{B(\xi)}, \qquad p(\eta)=\frac{\mu(\eta)-A(\eta)}{B(\eta)} 
\end{gather} 
with $A(z)$, $B(z)$, $C(z)$ given by (\ref{ABC}) and           
$\mu^{2}(z)=A^{2}(z)+B(z)C(z)$.  
 
\section{Explicit map and an equivalent approach to Darboux-dressing matrix} \label{sec3}
 
The matrix (\ref{Dexpl}) contains just one set of dynamical variables so that
the relation (\ref{eq:invpre}) gives now an explicit map between the variables
$\big(\tilde{s}^{+}_{j},\tilde{s}^{-}_{j},\tilde{s}^{3}_{j}\big)$ and
$\big(s^{+}_{j},s^{-}_{j},s^{3}_{j}\big)$. The map is easily found by (\ref{resj}); it reads: 
\begin{subequations} 
\begin{equation}\label{eq:s3t}\begin{split} 
\tilde{s}^{3}_{k}\, =& \,{\frac {p(\xi)p(\eta) \left( {\xi}^{2}-{\eta}^{2} \right)  \left( 
    \left( {z_{{k}}}^{2}-{\eta}^{2} \right) p(\xi)\xi- \left( {z_{{k}}}^{2}-{\xi}^{2} 
    \right) p(\eta)\eta \right) s^{-}_{k}z_{k}}{\Delta_{k}}}+\\ 
&{\frac { \left( {\xi}^{2}-{\eta}^{2} \right)  \left(  \left( {z_{{k}}}^{2}-{\xi}^{2} \right) p(\xi)\eta-p(\eta)\xi \left( {z_{{k}}}^{2}-{\eta}^{2} \right)  \right) 
\mbox{}{s}^{+}_{k}z_{k}}{\Delta_{k}}}+\\ 
&{\frac {s_{k}^{3}\Big[ p(\xi)p(\eta) \left(  \left( {\xi}^{2}+{z_{{k}}}^{2}
    \right)  \left( {\eta}^{2}+{z_{{k}}}^{2} \right) - \left(
    {\eta}^{2}+{\xi}^{2} \right)- 8{\eta}^{2}{\xi}^{2}{z_{{k}}}^{2}
    \right)}{\Delta_{k}}}+\\
&{-\frac{\left( \eta\xi \left( {\xi}^{2}-{z_{{k}}}^{2} \right)  \left( {\eta}^{2}-{z_{{k}}}^{2} \right)  \big( {p(\xi)}^{2}+{p(\eta)}^{2} \big)\mbox{} \right)\Big]}{{\Delta_{k}}}}
\end{split}\end{equation} 
 
\begin{equation}\label{eq:spt}\begin{split} 
\tilde{s}^{+}_{k}\, =& \,-{\frac {{b}^{2}{p(\xi)}^{2}{p(\eta)}^{2} \left( {\eta}^{2}-{\xi}^{2} \right) ^{2}s^{-}_{k}z_{k}^{2}}{\xi\eta\Delta_{k}}}+{\frac {{b}^{2} \left(  \left( {z_{{k}}}^{2}-{\xi}^{2} \right) p(\xi)\eta-p(\eta)\xi \left( {z_{{k}}}^{2}-{\eta}^{2} \right)  \right) ^{2} 
s^{+}_{k}}{\eta\xi\Delta_{k}}}+\\ 
&\,{\frac {2{b}^{2}p(\xi)p(\eta) \left( {\xi}^{2}-{\eta}^{2} \right)  \left(  \left( {z_{{k}}}^{2}-{\xi}^{2} \right) p(\xi)\eta-p(\eta)\xi \left( {z_{{k}}}^{2}-{\eta}^{2} \right)  \right)  
\mbox{}s^{3}_{k}z_{k}}{\eta\xi\Delta_{k}}} 
\end{split}\end{equation}
 
\begin{equation}\label{eq:smt}\begin{split} 
\tilde{s}^{-}_{k}\, =& \,-{\frac { \left( {\eta}^{2}-{\xi}^{2} \right) 
    ^{2}s^{+}_{k}z_{k}^2\xi\eta}{{b}^{2}\Delta_{k}}}+{\frac { \left(  \left({z_{{k}}}^{2}-{\eta}^{2} \right) p(\xi)\xi- \left( {z_{{k}}}^{2}-{\xi}^{2} \right) p(\eta)\eta \right) ^{2}s^{-}_{k}\xi\eta}{{b}^{2}\Delta_{k}}}+\\ 
&{\frac {2 \left( {\xi}^{2}-{\eta}^{2} \right)  \left(  \left( {z_{{k}}}^{2}-{\eta}^{2} \right) p(\xi)\xi- \left( {z_{{k}}}^{2}-{\xi}^{2} \right) p(\eta)\eta \right) s^{3}_{k}z_{k}\xi\eta}{{b}^{2}\Delta_{k}}} 
\end{split}\end{equation}\end{subequations} 
where $\Delta_{k}$ is proportional to the determinant of $D(z_{k})$, i.e.  
\begin{equation} 
\Delta_{k}=(z^{2}_{k}-\xi^{2})(z_{k}^{2}-\eta^{2})(p(\xi)\eta-p(\eta)\xi)(p(\eta)\eta-p(\xi)\xi) 
\end{equation} 
Formulas (\ref{eq:s3t}), (\ref{eq:spt}), (\ref{eq:smt}) define a two-parameter B\"acklund transformation, the parameters being $\xi$ and $\eta$: as we will show in the next section,  it is a crucial point to have a \emph{two}-parameter family of transformations when looking  for a physical map from real variables to real variables.  
As mentioned in the previous Section, we now show that indeed, by posing: 
\begin{equation} \label{bsubst} 
b= i\sqrt{\eta\xi} 
\end{equation} 
the expression (\ref{Dexpl}) of the dressing matrix goes into the expression of the elementary Lax matrix for the classical, partially anisotropic, Heisenberg spin chain on the lattice \cite{FT}. 

\noindent 
Obviously two matrices differing only for a global multiplicative factor give rise to the same similarity transformation. So we omit the term $\frac{\beta z}{z^{2}-\xi^{2}}$ in (\ref{Dexpl}), and, taking into account (\ref{bsubst}), we write for the diagonal part $D_{d}$ of (\ref{Dexpl}): 
\begin{equation} \label{Dd} 
D_{d} = \frac{i}{2}\Big((p(\xi)-p(\eta))(v - w)\dblone+(p(\xi)+p(\eta))(v+w)\sigma_{3}\Big) 
\end{equation} 
where $v (\xi, \eta)$ and $w (\xi,\eta)$ are given by: 
\begin{equation} 
v (\xi,\eta)=\frac{z\xi}{\sqrt{\eta\xi}}-\frac{\eta \sqrt{\eta\xi}}{z} \qquad w(\xi,\eta)=\frac{\xi\sqrt{\eta\xi}}{z}-\frac{z\eta}{\sqrt{\eta\xi}} =  -v (\eta,\xi)
\end{equation} 
We substitute: 
\begin{equation} \label{eq33}
\xi\to e^{i\zeta_{1}} \qquad \eta \to e^{i\zeta_{2}} \qquad z\to e^{i\lm}  
\end{equation} 
and take a suitable redefinition of the B\"acklund parameters to clarify the
structure of the $D$ matrix:
\begin{equation}\label{lmu}
\lambda_{0}\doteq\frac{\zeta_{1}+\zeta_{2}}{2} \qquad \mu \doteq \frac{\zeta_{1}-\zeta_{2}}{2}
\end{equation}
With these positions it is simple to find that
$v-w=4ie^{i\lambda_{0}}\sin(\lambda-\lambda_{0})\cos(\mu)$ and
$v+w=4ie^{i\lambda_{0}}\cos(\lambda-\lambda_{0})\sin(\mu)$. So, considering
equation (\ref{Dd}) jointly with the off-diagonal part of (\ref{Dexpl}), the dressing matrix can be written as: 
 
\begin{equation} \label{eq:Dprov}\begin{split}
D(\lambda)=&\alpha
\Big[\sin(\lambda-\lambda_{0})\dblone+\frac{p(\zeta_{1})+p(\zeta_{2})}{p(\zeta_{1})-p(\zeta_{2})}\tan(\mu)\cos(\lambda-\lambda_{0})\sigma_{3}+\\
&+\frac{2\sin(\mu)}{p(\zeta_{2})-p(\zeta_{1})}\left(\begin{array}{cc}0 & 1\vspace{1mm}\\ 
-p(\zeta_{1})p(\zeta_{2}) & 0\end{array}\right)\Big] 
\end{split}\end{equation} 
where $\alpha$ is the global factor
$2e^{i\lambda_{0}}(p(\zeta_{2})-p(\zeta_{1})) $. Observe that in formula
(\ref{eq:Dprov}), with some abuse of notation, $p(\zeta_{1})$
$\left(p(\zeta_{2})\right)$ stands of course for
$\left.p(\xi)\right|_{\xi=e^{i\zeta_{1}}}$
$\left(\left.p(\eta)\right|_{\eta=e^{i\zeta_{2}}}\right)$. \newline A last change of variables allows to identify the dressing matrix with the elementary Lax matrix of the classical $xxz$ Heisenberg spin chain on the lattice, and furthermore to recover the form of the Darboux matrix for the \emph{rational} Gaudin model \cite{HKR}\cite{SK1} in the limit of \emph{small angles}. Namely, we introduce two new functions, $P$ and $Q$, by letting
\begin{equation} 
p(\zeta_{1})=-Q \qquad p(\zeta_{2}) = \frac{2\sin(\mu)}{P}-Q.  
\end{equation} 
Then equation (\ref{eq:Dprov}) becomes:  
\begin{equation} \label{eq:Darboux} 
D(\lambda)=\alpha \left(\begin{array}{cc} \sin(\lambda-\lambda_{0}-\mu)+PQ\,\cos(\lambda-\lambda_{0}) & 
  P\,\cos(\mu)\\ 
Q\,\sin(2\mu)-PQ^{2}\cos(\mu) & \sin(\lambda-\lambda_{0}+\mu)-PQ\,\cos(\lambda-\lambda_{0}) 
\end{array}\right) 
\end{equation} 
Obviously now we can repeat the argument made before about spectrality; indeed now $D\big|_{\lambda=\lm_{0}+\mu}$ and $D\big|_{\lambda=\lm_{0}-\mu}$ are rank one matrices. So if $\Omega_{+}$ and $\Omega_{-}$ are respectively the kernels of $D(\lambda_{0}+\mu)$ and $D(\lambda_{0}-\mu)$ one has again that   
$\Omega_{+}$ and $\Omega_{-}$ are eigenvectors of  $L(\lambda_{0}+\mu)$ 
and $L(\lambda_{0}-\mu)$ with eigenvalues $\gamma_{+}$ and $\gamma_{-}$  where
$$ 
\gamma_{\pm}=\gamma(\lambda)\Big\|_{\lambda=\lambda_{0}\pm\mu}$$
\noindent
and we have set (\ref{generfun})
\begin{equation}\label{gammas}
\gamma^{2}(\lambda) \equiv  A^{2}(\lambda)+B(\lambda)C(\lambda)=-\textrm{det}(L(\lambda))
\end{equation}
The two kernels are given by: 
\begin{equation} 
\Omega_{+}=\left(\begin{array}{c}1\\-Q\end{array}\right) \qquad \Omega_{-}=\left(\begin{array}{c}P\\2\,\sin(\mu)-PQ\end{array}\right) 
\end{equation} 
and the eigenvectors relations yields the following expression  of $P$ and $Q$ in terms of the old variables only: 
\begin{equation}\label{eq:P&Q} 
Q=Q(\lambda_{0}+\mu)=\frac{A(\lambda)-\gm(\lambda)}{B(\lambda)}\Big\|_{\lambda=\lambda_{0}+\mu} \qquad \frac{1}{P}=\frac{Q(\lambda_{0}+\mu)-Q(\lambda_{0}-\mu)}{2\,\sin(\mu)} 
\end{equation} 
The explicit map can be found by equating the residues at the poles
$\lambda=\lambda_{k}$ in (\ref{BTnew}), that is by the relation:   
\begin{equation} \label{resk}
\tl{L}_{k}D_{k}=D_{k}L_{k} 
\end{equation} 
where 
\begin{equation}  
L_{k}=\left(\begin{array}{cc}s^{3}_{k} & s^{-}_{k}\\ s^{+}_{k} & 
  -s^{3}_{k}\end{array}\right),\qquad D_{k}=D(\lambda=\lambda_{k}) 
\end{equation} 
or by performing the needed changes of variables in (\ref{eq:s3t}), (\ref{eq:spt}), (\ref{eq:smt}). Anyway now the map reads: 
\begin{subequations}\begin{equation}
\begin{split} \label{eqs3}
\tilde{s}^{3}_{k}=&\frac{2\,\cos^{2}(\mu)-(\cos^{2}(\mu)+\cos^{2}(\delta^{k}_{0}))(1-2PQ\,\sin(\mu)+P^{2}Q^{2})}{\Delta_{k}}s^{3}_{k}+\\ 
&+\frac{P\,\cos(\mu)(\sin(\delta^{k}_{+})-PQ\,\cos(\delta^{k}_{0}))}{\Delta_{k}}s^{+}_{k}+\\
&-\frac{Q\,\cos(\mu)(2\,\sin(\mu)-PQ)(\sin(\delta^{k}_{-})+PQ\,\cos(\delta^{k}_{0}))}{\Delta_{k}}s^{-}_{k} 
\end{split}\end{equation} 
\begin{equation}\begin{split}  \label{eqsp}
\tilde{s}^{+}_{k}&=\frac{(\sin(\delta_{+}^{k})-PQ\,\cos(\delta_{0}^{k}))^{2}}{\Delta_{k}}s^{+}_{k}-\frac{(Q^{2}\cos^{2}(\mu)(2\,\sin(\mu)-PQ))^{2}}{\Delta_{k}}s^{-}_{k}+\\ 
&+\frac{2Q\,\cos(\mu)(2\,\sin(\mu)-PQ)(\sin(\delta_{+}^{k})-PQ\,\cos(\delta^{k}_{0}))}{\Delta_{k}}s^{3}_{k} 
\end{split}\end{equation} 
\begin{equation}\begin{split}\label{eqsm} 
\tilde{s}_{k}^{-}=&\frac{(\sin(\delta_{-}^{k})+PQ\,\cos(\delta_{0}^{k}))^{2}}{\Delta_{k}}s_{k}^{-}-\frac{P^{2}\,\cos^{2}(\mu)}{\Delta_{k}}s_{k}^{+}+\\
&-\frac{2P\,\cos(\mu)(\sin(\delta^{k}_{-})+PQ\,\cos(\delta_{0}^{k}))}{\Delta_{k}}s_{k}^{3} 
\end{split}\end{equation}\end{subequations} 
where for typesetting brevity we have put: 
\begin{equation}\left\{\begin{array}{ll} 
\delta_{0}^{k}=\lambda_{k}-\lm_{0}\\ 
\delta_{\pm}^{k}=\lambda_{k}-\lm_{0}\pm \mu 
\end{array}\right. 
\end{equation} 
and we have denoted by $\Delta_{k}$ the determinant of $D(\lambda_{k})$, that
is: 
$$ 
\Delta_{k}:=\sin(\lambda_{k}-\lambda_{0}-\mu)\sin(\lambda_{k}-\lambda_{0}+\mu)(1-2PQ\,\sin(\mu)+P^{2}Q^{2})$$ 
At this point we can show that for ``small'' $\lambda_{0}$ and $\mu$ 
one obtains, at first order, the B\"acklund for the rational 
Gaudin model, independently found by Sklyanin \cite{SK1} on one hand and Hone, Kuznetsov and Ragnisco \cite{HKR} on the other, as the composition of two one-parameter B\"acklunds.    
So let us take $\lambda_{0} \to h\lambda_{0}$, $\mu \to h\mu$ and $\lambda \to h\lambda$ where $h$ is the expansion parameter. One has: 
$$ 
\cot(\lambda-\lambda_{k})=\frac{1}{h(\lambda-\lambda_{k})}+O(h)\qquad \frac{1}{\sin(\lambda-\lambda_{k})}=\frac{1}{h(\lambda-\lambda_{k})}+O(h), 
$$ 
so that $Q = q^{r}+O(h^{2})$, 
where the superscript $r$ stands for ``rational'' . Thus, $q^{r}$ coincides with 
the variable $q$ that one finds in the rational case \cite{HKR}. For the variable $P$ one has:  
$$ 
P = h(p^{r}+O(h^{2})) \qquad \textrm{where} \qquad p^{r}=\frac{2\mu}{q^{r}(\lambda_{0}+\mu)-q^{r}(\lambda_{0}-\mu)} 
$$ 
Taking into account these expressions, it is straightforward to see that the
matrix (\ref{eq:Darboux}) has the expansion:
\begin{equation}
D(\lambda)=hD^{r}(\lambda)+O(h^{3}) 
\end{equation}
where 
\begin{equation} D^{r}(\lambda)= \left(\begin{array}{cc} \lambda-\lm_{0}-\mu +p^{r}q^{r} & p^{r}\\  
q^{r}(2\mu-p^{r}q^{r}) & \lambda-\lm_{0}+\mu-p^{r}q^{r} \end{array}\right). 
\end{equation} 
The limit of ``small angles'' in (\ref{eq:s3t}), (\ref{eq:spt}),
(\ref{eq:smt}) obviously leads to the rational map of \cite{HKR}.
\subsection{Symplecticity}    
In this subsection we face  the question of  the simplecticity of our map; the correspondence with the rational B\"acklund in the limit of ``small angles'' shows that the transformations are surely canonical in this limit. Indeed,  as our map is explicit, we could check by brute-foce calculations  whether the Poisson structure (\ref{poisS}) is preserved by tilded variables.  However  we will follow a finer argument due to Sklyanin \cite{SK2}.  
Suppose that $D(\lambda)$ obeys the \emph{quadratic} Poisson bracket, that is 
\begin{equation}\label{eq:PoisD} 
\{D^{1}(\lambda),D^{2}(\tau)\}=[r_t(\lambda-\tau),D^{1}(\lambda)\otimes D^{2}(\tau)] 
\end{equation}
where as usually $D^{1}=D\otimes\dblone$, $D^{2}=\dblone\otimes D$. Consider the relation 
\begin{equation}\label{eq:transf} 
\tilde{L(\lambda)}\tilde{D}(\lambda-\lambda_{0})=D(\lambda-\lambda_{0})L(\lambda) 
\end{equation} 
in an extended phase space, where the entries of $D$ Poisson commutes with those of $L$. 
Note that in (\ref{eq:transf}) we have used tilded variables also for $D(\lambda)$  (in its l.h.s.) because (\ref{eq:transf}) is indeed  the B\"acklund transformation  in this  extended phase space, whose coordinates are $(s^{3}_{j},s^{\pm}_{j},P,Q)$,  so that we have also a $\tl{P}$ and a $\tl{Q}$. The key observation is that if both $L$ and $D$ have the 
same Poisson structure, given by equation (\ref{eq:PoisD}), then this property
holds true for $LD$ and 
$DL$ as well, because in this extended space the 
entries of $D$ Poisson commute with the entries of $L$. This means that the 
transformation (\ref{eq:transf}) defines a ``canonical'' 
transformation. Sklyanin showed \cite{SK2} that if one now restricts the variables on the constraint manifold $\tl{P}=P$ and $\tl{Q}=Q$ the symplecticity is 
preserved; however this constraint leads to a dependence of $P$ and $Q$ on 
the entries of $L$, that for consistency must be the same as the one   given by the 
equation (\ref{eq:transf}) on this constrained manifold. But there 
(\ref{eq:transf}) is just given by the usual BT: 
$$ 
\tilde{L}(\lambda)D(\lambda-\lambda_{0})=D(\lambda-\lambda_{0})L(\lambda) 
$$       
so that the map  preserves the spectrum of $L(\lambda)$ and is canonical. 
What remains to show is that indeed 
(\ref{eq:PoisD}) is fullfilled by our $D(\lambda)$.  
Obviously $D(\lambda)$ cannot have this Poisson structure for any Poisson
bracket between $P$ and $Q$. In the rational case the Darboux matrix has the Poisson structure imposed by the rational $r$-matrix provided $P$ and $Q$ are 
canonically conjugated in the extended space \cite{SK2} (and this is why they
were called $P$ and $Q$); in the trigonometric case $P$ and $Q$ are no longer
canonically conjugated but obviously one recovers this property at order $h$ in the ``small angle'' limit. 

\noindent 
First note that $D(\lambda)$ can be conveniently written as: 
\begin{equation}\label{eq:Dlam} 
D(\lambda) = \alpha\,\cos(\mu)\Big[\sin(\lambda)\dblone+a\,\cos(\lambda)\sigma_{3}+\left(\begin{array}{cc}0 & b\vspace{1mm}\\ 
c & 0\end{array}\right)\Big] 
\end{equation} 
where the coefficients $a, b, c$ are given by: 
\begin{equation} 
a=\frac{PQ-\sin(\mu)}{\cos(\mu)}, \quad b=P, \quad c=2Q\,\sin(\mu)-PQ^{2} 
\end{equation} 
Inserting (\ref{eq:Dlam}) in (\ref{eq:PoisD}) we have the following constraints: 
\begin{equation} \label{eq:pois1} 
\{\alpha,\alpha a \}=0  \quad \Longrightarrow \quad \alpha=\alpha(PQ) 
\end{equation} 
\begin{equation}\label{eq:pois2} 
\{\alpha,\alpha b \}=-\alpha^{2}ab  \quad \Longrightarrow \quad 
\{\alpha,P\}=\alpha P\frac{\sin(\mu)-PQ}{\cos(\mu)} 
\end{equation} 
\begin{equation}\label{eq:pois3} 
\{\alpha,\alpha c \}=\alpha^{2}ac  \quad \Longrightarrow \quad \{\alpha,Q\}=-\alpha Q\frac{\sin(\mu)-PQ}{\cos(\mu)} 
\end{equation} 
All remaining relations, namely 
\begin{equation}\label{eq:pois456} 
\{\alpha b,\alpha c\}=2\alpha^{2}a \quad \{\alpha a,\alpha b\}=\alpha^{2}b 
\quad \{\alpha a,\alpha c\}=-\alpha^{2}c 
\end{equation}  
give the same constraint, i.e.: 
\begin{equation}\label{eq:poispq} 
\{Q,P\}=\frac{1+P^{2}Q^{2}-2PQ\sin(\mu)}{\cos(\mu)} 
\end{equation} 
This expression can be used to find, after a simple integration, 
$$ 
\alpha(PQ)=\frac{k}{\sqrt{(1+P^{2}Q^{2}-2PQ\sin(\mu))}} 
$$ 
so that the Darboux matrix (\ref{eq:Darboux}) is fixed (up to the constant 
multiplicative factor $k$). 
As previously pointed out, it turns out that    the Darboux-dressing matrix (\ref{eq:Darboux}) is formally equivalent to the elementary Lax matrix for the classical $xxz$ Heisenberg spin chain on the lattice \cite{FT}. Moreover it has also the same (quadratic) Poisson bracket. This suggests that indeed $D(\lambda)$ can be recast in the form (see \cite{FT}): 
\begin{equation}\label{eq:DdiFT} 
D(\lambda)=\mathscr{S}_{0}1+\frac{i}{\sin(\lambda)}\big(\mathscr{S}_{1}\sigma_{1}+\mathscr{S}_{2}\sigma_{2}+\cos(\lambda)\mathscr{S}_{3}\sigma_{3}\big) 
\end{equation}  
where the $\sigma_{i}$ are the Pauli matrices and the variables
$\mathscr{S}_{i}$ satisfies the following Poisson bracket (\cite{FT}): 
\begin{equation}\label{eq:PoisDdiFT} \begin{array}{cc} 
\{\mathscr{S}_{i},\mathscr{S}_{0}\}=J_{jk}\mathscr{S}_{j}\mathscr{S}_{k}\\ 
\{\mathscr{S}_{i},\mathscr{S}_{j}\}=-\mathscr{S}_{0}\mathscr{S}_{k} \end{array} 
\end{equation} 
where $(i,j,k)$ is a cyclic permutation of $(1,2,3)$ and $J_{jk}$ is 
antisymmetric with $J_{12}=0, J_{13}=J_{23}=1$. Indeed it is 
straightforward to show that the link between the two representations 
(\ref{eq:Dlam}) and (\ref{eq:DdiFT}), up to the factor
$\cos(\mu)\sin(\lambda)$ that does not affect neither (\ref{BTnew}) nor the Poisson bracket (\ref{eq:PoisD}), is given by : 
\begin{equation} 
\alpha=\mathscr{S}_{0} \quad -\frac{i\alpha}{2}(b+c)=\mathscr{S}_{1} \quad 
\frac{\alpha}{2}(b-c)=\mathscr{S}_{2} \quad -ia\alpha=\mathscr{S}_{3} 
\end{equation} 
and the Poisson brackets (\ref{eq:pois1}), (\ref{eq:pois2}), (\ref{eq:pois3}), 
(\ref{eq:pois456}) correspond to those given in (\ref{eq:PoisDdiFT}).
\newline  
An open question regards the generating function of our BT. So far we have not been able to write it in any closed form; in our opinion the question
is harder than in  the rational case (where the generating function is
known from \cite{HKR}): in fact the rational map corresponding to
(\ref{eq:s3t}), (\ref{eq:spt}), (\ref{eq:smt}) can be written as the
composition of two simpler \emph{one}-parameter B\"acklund transformations,
and this entails the same property to hold for the generating function; in the
trigonometric case a factorization of the B\"acklund transformations cannot
preserve the symmetry (\ref{Symm}): so probably one should look for symmetry-violating generating functions such that their composition enables symmetry to be restored.  
\section{Physical B\"acklund transformations}\label{sec4} 
The transformations we have found do not map, in general, real variables into
real variables. A sufficient condition to ensure this property is given by: 
\begin{equation}
\zeta_{1}=\bar{\zeta}_{2}\label{reality}
\end{equation}
which amounts te require that  $\lambda_{0}$ and $\mu$ in (\ref{eqs3}), (\ref{eqsp}),
(\ref{eqsm})  be, respectively,  real and  imaginary
numbers. 
\newline
Indeed we claim that, if (\ref{reality}) holds,  starting from a physical solution of the dynamical equations, we
can find a new physical solution with two real parameters. Let us prove the
assertion. B\"acklund transformation are obtained by (\ref{resk});
starting from  a real solution means starting from an Hermitian $L_{k}$. Thus,  if the transformed matrix $\tl{L}_{k}$ has to be Hermitian
too,  the Darboux matrix has to be proportional to an unitary matrix. We will
 show that this is the case by choosing $\zeta_{1}=\bar{\zeta}_{2}$ and
$\gamma(\zeta_{1})=-\bar{\gamma}(\zeta_{2})$ ($\gamma$ is the function defined
 in (\ref{gammas})). Note that the condition on the $\gamma$'s specifies their
relative sign (the sheet on the Riemann surface), inessential for the
spectrality property. Hereafter we assume the parameter $\mu$, defined in
(\ref{eq33}), to be purely imaginary $=i\epsilon$, so that: 
\begin{equation}\label{epsmu}
\zeta_{1}=\lambda_{0}+i\epsilon  \qquad (\lm_{0},\epsilon)\in\mathbb{R}^{2}
\end{equation}
The Darboux matrix at $\lambda=\lambda_{k}$ can be rewritten as:
\begin{equation}\label{eq:Dar1}
D_{k}=\left(\begin{array}{cc}
 \sin(v_{k}-i\epsilon)+PQ\,\cos(v_{k}) & P\,\cosh(\epsilon)\\
 Q\,\cosh(\epsilon)\,(2i\,\sinh(\epsilon)-PQ) & \sin(v_{k}+i\epsilon)-PQ\,\cos(v_{k})\end{array}\right)
\end{equation}
where $v_{k} \equiv \lambda_{k}-\lambda_{0}$ (we are assuming that the
parameters $\lambda_{k}$ of the model are real). 
We recall that in (\ref{eq:Dar1}):
\begin{equation} \label{eq:peq}
Q=Q(\zeta_{1})=\frac{A(\zeta_{1})-\gm(\zeta_{1})}{B(\zeta_{1})}=-\frac{C(\zeta_{1})}{A(\zeta_{1})+\gamma(\zeta_{1})}; \qquad
P=\frac{2i\,\sinh(\epsilon)}{Q(\zeta_{1})-Q(\bar{\zeta_{1}})}.
\end{equation} 
Furthermore it is a  simple matter to show that 
\begin{equation}\label{eq:ABC}
A(\zeta_{1})=\bar{A}(\bar{\zeta_{1}}); \quad B(\zeta_{1})=\bar{C}(\bar{\zeta_{1}}); \quad C(\zeta_{1})=\bar{B}(\bar{\zeta_{1}}).
\end{equation}
If the off-diagonal terms of $D_{k}D_{k}^{\dag}$ has to be zero, then the
following equation has to be fullfilled:
\begin{equation} \label{eq:firsteq}
P(\sin(v_{k}-i\epsilon)-\bar{P}\bar{Q}\,\cos(v_{k}))=\bar{Q}(2i\,\sinh(\epsilon)+\bar{P}\bar{Q})(\sin(v_{k}-i\epsilon)+PQ\,\cos(v_{k}))
\end{equation}
Using relations (\ref{eq:peq}) and rearranging the terms, the previous equation becomes: 
\begin{equation}\label{eq:secondeq}\begin{split}
&(\frac{1}{\bar{Q}(\zeta_{1})}-\frac{1}{\bar{Q}(\bar{\zeta_{1}})})\cosh(\epsilon)\sin(v_{k})+i(\frac{1}{\bar{Q}(\zeta_{1})}+\frac{1}{\bar{Q}(\bar{\zeta_{1}})})\cos(v_{k})\sinh(\epsilon)=\\
&=(Q(\zeta_{1})-Q(\bar{\zeta_{1}}))\cosh(\epsilon)\sin(v_{k})+i\cos(v_{k})\sinh(\epsilon)(Q(\zeta_{1})-Q(\bar{\zeta_{1}}))
\end{split}\end{equation}
Note that the relations (\ref{eq:ABC}) gives
$\gamma^{2}(\zeta_{1})=\overline{\gamma^{2}(\bar{\zeta_{1}})}$, implying that
$\gamma^{2}(\lambda)$ is a real function of its complex argument, consistently
with the expansion (\ref{generfun}). \newline The choice:
\begin{equation}\label{gamma12}
\gamma(\zeta_{1})=-\bar{\gamma}(\bar{\zeta_{1}})
\end{equation}
entails:
\begin{equation}
\bar{Q}(\zeta_{1})=-\frac{1}{Q(\bar{\zeta_{1}})}
\end{equation}
With this constraint the equation
(\ref{eq:secondeq}) holds too. Moreover (\ref{gamma12})  makes the diagonal
terms in $D_{k}D_{k}^{\dag}$ equal. This shows that, under the given
assumptions, $D_{k}$ is an unitary matrix.  
 
\section{Interpolating Hamiltonian flow} \label{sec5}
The B\"acklund transformation can be seen as a time discretization of a
one-parameter ($\lambda_{0}$) family of hamiltonian flows with the difference
$i(\bar{\zeta_{1}}-\zeta_{1})=2\epsilon$ playing the role of the time-step. To clarify this point, let
us take the limit $\epsilon \to 0$.\\
We have:
\begin{equation}\label{eq:q0}
Q=\frac{A(\lambda_{0})-\gamma(\lambda_{0})}{B(\lambda_{0})}+O(\epsilon)\equiv Q_{0}+O(\epsilon)
\end{equation}
\begin{equation}\label{eq:p0}
P=-i\epsilon\frac{B(\lambda_{0})}{\gamma(\lambda_{0})}+O(\epsilon^{2})\equiv
i\epsilon P_{0}+O(\epsilon^{2})
\end{equation}
and for the dressing matrix we can write:
\begin{equation}\begin{split}
&D(\lambda)=k\,\sin(\lambda-\lambda_{0})\dblone+\\
+&i\epsilon k\left(\begin{array}{cc}
\cos(\lambda-\lambda_{0})(P_{0}Q_{0}-1) & P_{0}\\
Q_{0}(2-P_{0}Q_{0})& \cos(\lambda-\lambda_{0})(1-P_{0}Q_{0})
\end{array}\right)+O(\epsilon^{2})\end{split}
\end{equation}
Reorganizing the terms with the help of $P_{0}$ and $Q_{0}$ given in the equations (\ref{eq:q0}) and (\ref{eq:p0}) we
arrive at the expression:
\begin{equation}\begin{split}
&D(\lambda)=k\,\sin(\lambda-\lambda_{0})\dblone+\\
-&\frac{i\epsilon k}{\gamma(\lambda_{0})}\left(\begin{array}{cc}
A(\lambda_{0})\cos(\lambda-\lambda_{0}) & B(\lambda_{0})\\
C(\lambda_{0})& -A(\lambda_{0})\cos(\lambda-\lambda_{0})
\end{array}\right)+O(\epsilon^{2})\end{split}
\end{equation}
It is now straightforward  to show that in the  limit $\epsilon \to 0$ the equation of the map
$\tl{L}D=DL$ turns into the Lax equation for a continuous flow: 
\begin{equation}\label{eq:motion}
\dot{L}(\lm)= [L(\lambda),M(\lambda,\lambda_{0})]
\end{equation}
where the time derivative is defined as:
\begin{equation}
\dot{L}=\lim_{\epsilon\rightarrow 0}\frac{\tilde{L}-L}{\epsilon}
\end{equation}
and the matrix $M(\lambda,\lambda_{0})$ has the form
\begin{equation}
\frac{i}{\gamma(\lambda_{0})}\left(\begin{array}{cc}
A(\lambda_{0})\cot(\lambda-\lambda_{0}) & \frac{B(\lambda_{0})}{\sin(\lambda-\lambda_{0})}\\
\frac{C(\lambda_{0})}{\sin(\lambda-\lambda_{0})}& -A(\lambda_{0})\cot(\lambda-\lambda_{0})
\end{array}\right)
\end{equation}
The system (\ref{eq:motion})
can be cast in Hamiltonian  form:
\begin{equation}
\dot{L}(\lambda)=\{\mathcal{H}(\lambda_{0}),L(\lambda)\}
\end{equation} 
with the Hamilton's function given by:
\begin{equation}\label{eq:Ham}
\mathcal{H}(\lambda_{0})=\gamma(\lambda_{0})=\sqrt{A^{2}(\lambda_{0})+B(\lambda_{0})C(\lambda_{0})}
\end{equation}
Quite remarkably, but not surprisingly, the Hamiltonian (\ref{eq:Ham})
characterizing the interpolating flow is (the square root of) the generating
function (\ref{generfun}) of the whole set of conserved quantities. By
choosing the parameter $\lambda_{0}$ to be equal to any of the poles ($\lambda_{i}$) of the
Lax matrix, the map leads to $N$ different maps $\{BT^{(i)}\}_{i=1..N}$, where
$BT^{(i)}$ discretizes the flow
corresponding to the Hamiltonian $H_{i}$, given by equation
(\ref{hams}). Any other integrable map for the trigonometric Gaudin model can be, in principle, written in terms of the
$N$ maps $\{BT^{(i)}\}_{i=1..N}$. \newline
More explicitely, by posing $\lambda_{0}=\delta +\lambda_{i}$ and taking
the limit $\delta \to 0$, the Hamilton's function (\ref{eq:Ham}) gives:
\begin{equation}
\gamma(\lambda_{0})=\frac{s_{i}}{\delta}+\frac{H_{i}}{2s_{i}}+O(\delta)
\end{equation}
and the equations of motion take the form:
\begin{equation}
\dot{L}(\lambda)=\frac{1}{2s_{i}}\{H_{i},L(\lambda)\}
\end{equation}
Accordingly, the interpolating flow encompasses all the commuting flows of
the system, so that the B\"acklund transformations turn out to be an
\emph{exact time-discretizations} of such interpolating flow.   
\subsection{Numerics}
\begin{figure}[ht]
\centering
\includegraphics[scale=0.4]{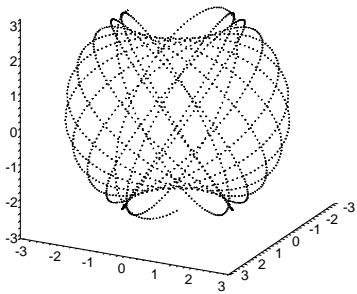}
\caption{input parameters: $s_{1}^{+}=2+i$, $s_{1}^{-}=2-i$, $s_{1}^{3}=-2$,
$s_{2}^{+}=50+40i$, $s_{2}^{-}=50-40i$, $s_{2}^{3}=70$, $\lambda_{1}=\pi/110$,
$\lambda_{2}=7\pi/3$, $\lambda_{0}=0.1$, $\mu=-0.002i$}
\end{figure}
\begin{figure}[ht]
\centering
\includegraphics[scale=0.4]{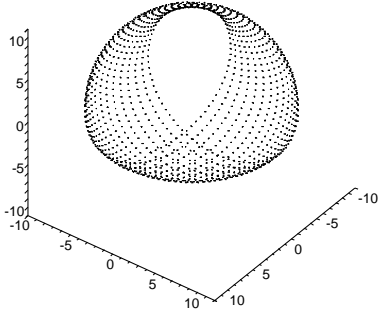}
\caption{input parameters: $s_{1}^{+}=0.2+10i$, $s_{1}^{-}=0.2-10i$, $s_{1}^{3}=-1$,
$s_{2}^{+}=10-30i$, $s_{2}^{-}=10+30i$, $s_{2}^{3}=100$, $\lambda_{1}=\pi$,
$\lambda_{2}=7\pi/3$, $\lambda_{0}=0.1$, $\mu=-0.004i$}
\end{figure}
The figures report an example of iteration of the map (\ref{eqs3}),
(\ref{eqsp}), (\ref{eqsm}). For simplicity we take $N=2$. The computations
shows the first $1500$ iterations: the plotted variables are the
physical ones $(s_{1}^{x},s_{1}^{y},s_{1}^{z})$. Only one of the two spins is
shown, namely that labeled by the subscript ``1''. The figures are obtained by a \texttrademark Maple\, code.

\end{document}